\documentclass[reprint,amsmath,amssymb,aps,showpacs,floatfix,superscriptaddress]{revtex4-1}
\pdfoutput=1
\usepackage{times,color,graphicx}
\usepackage[colorlinks=true,urlcolor=blue,linkcolor=blue,citecolor=blue]{hyperref}

\usepackage{natbib}
\usepackage[colorlinks=true,linkcolor=blue,citecolor=blue]{hyperref}%
\usepackage{MnSymbol}%
\usepackage{wasysym}%
\usepackage{epstopdf}

\begin{document}

\title{Characterisation of p-type ZnS:Cu transparent conducting films fabricated by \\high-temperature pulsed laser deposition}

\author{K. S. Duncan}
 \affiliation{School of Physics, HH Wills Physics Laboratory, University of Bristol, Tyndall Avenue, Bristol, BS8 1TL United Kingdom}
 
\author{J. D. Taylor}
 \affiliation{School of Physics, HH Wills Physics Laboratory, University of Bristol, Tyndall Avenue, Bristol, BS8 1TL United Kingdom}
 \affiliation{Department of Physics, University of Bath, Bath BA2 7AY, United Kingdom}

\author{M. Jonak}
 \affiliation{School of Physics, HH Wills Physics Laboratory, University of Bristol, Tyndall Avenue, Bristol, BS8 1TL United Kingdom}

\author{K. O. E. Derricutt}
 \affiliation{School of Physics, HH Wills Physics Laboratory, University of Bristol, Tyndall Avenue, Bristol, BS8 1TL United Kingdom}

\author{\\A. G. J. Tallon}
 \affiliation{School of Physics, HH Wills Physics Laboratory, University of Bristol, Tyndall Avenue, Bristol, BS8 1TL United Kingdom}

\author{C. E. Wilshaw}
 \affiliation{School of Physics, HH Wills Physics Laboratory, University of Bristol, Tyndall Avenue, Bristol, BS8 1TL United Kingdom}

\author{J. A. Smith}
 \affiliation{School of Chemistry, University of Bristol, Bristol BS8 1TS, United Kingdom}

\author{N. A. Fox}
 \email{Neil.Fox@bristol.ac.uk}
 \affiliation{School of Physics, HH Wills Physics Laboratory, University of Bristol, Tyndall Avenue, Bristol, BS8 1TL United Kingdom}
 \affiliation{School of Chemistry, University of Bristol, Bristol BS8 1TS, United Kingdom}
 
\begin{abstract}
Copper-doped zinc sulphide (ZnS:Cu) thin films were synthesized through pulsed laser ablation in an inert background gas on stationary and rotating substrates, and a comprehensive opto-electrical characterisation is presented. The Cu$_x$Zn$_{1-x}$S films demonstrated comparable conductivity and transparency to other leading p-type transparent conducting materials, with a peak conductivity of 49.0 Scm$^{-1}$ and a hole mobility of 1.22 cm$^2$V$^{-1}$s$^{-1}$ for films alloyed with an x = 0.33 copper content. The most conducting films displayed a transparency of 71.8$\%$ over the visible range at a thickness of 100 nm, and band gaps were found in the range 3.22-3.52 eV, which showed a strong negative correlation with copper content. The effects of sulphur-rich rapid thermal annealing on the synthesized compound are reported, with films reliably displaying an increase in conductivity and carrier mobility. Films grown using a stationary substrate possessed large spatial thickness distributions and displayed sub-band gap absorption, which is discussed with respect to inhomogeneous copper substitution. Films deposited at 450$^\circ$C were found to be in the zincblende phase before and after annealing, with no occurrence of a phase change to wurtzite structure.   
\end{abstract}

\maketitle

\section{Introduction}
\label{introduction}
\indent
Transparent conducting materials (TCMs) are wide band gap semiconducting materials with high optical transparencies and low resistivities. TCMs play a central role within opto-electronics, and have passive uses as front contacts in LCD displays, as components in LEDs, and as window layers of solar panels \cite{wang2012effective,berginski2007effect}.

The most prevalent TCM, indium tin oxide (ITO), has 92.1$\%$ optical transparency, and average conductivities of 1400 Scm$^{-1}$, but is increasingly expensive to fabricate due to the scarcity of indium \cite{chen2013fabrication}. Various sustainable alternatives to ITO have been developed \cite{minami2008present,hecht2011emerging}. The development of efficient TCMs is also being driven by the need to harness renewable energy sources effectively. Structures that could utilise TCMs, such as heterojunctions, metal-insulator-semiconductor (MIS), or semiconductor-insulator-semiconductor (SIS) structures, would enable the fabrication of low-cost, efficient, and transparent photovoltaic devices.

Most TCMs, such as ITO, are n-type, thus comparatively less literature exists on p-type TCMs. The first reported instance of a p-type conductivity in a TCM was in delafossite CuAlO$_2$ which displayed hole conductivities of up to 1 Scm$^{-1}$ and an optical transmission of 70$\%$ \cite{kawazoe1997p}.  More recent work has found that among the best performing p-type materials are CuAlS$_2$ with a transmittance of 80$\%$ in the visible range and conductivity of 63.5 Scm$^{-1}$ \cite{liu2007p}, and BaCu$_2$S$_2$ with a transparency of up to 90$\%$ and a conductivity of 17 Scm$^{-1}$ \cite{wang2007solution}. Many n-type TCMs significantly outperform existing p-type TCMs, and so fewer p-type TCMs have been developed for commercial use. One reason for the higher performance of n-type TCMs is the higher mobility of electrons compared to that of holes. The hole effective mass of ZnS, however, is 0.61 m$_0$, a value which is relatively low compared to that of most metal oxides \cite{uzar2011synthesis,li2012native}, and as such should not constitute an obstacle to achieving high mobilities. 

ZnS is a weakly n-type wide band gap (3.6-3.9 eV) II-VI semiconductor, and exists in either zincblende or wurtzite phase. Both phases are of interest due to their UV range band gaps, ideal for transparent opto-electronic applications. Doping with Cu is one method of fabricating a ZnS-based p-type alloy, whilst preserving optical transparency. Valency considerations of Cu$^{1+}$ and Zn$^{2+}$ suggest that where dopant Cu replaces Zn on lattice sites, it should act as an acceptor. The alloy is expected to remain n-type for low Cu contents, and to become p-type at a critical percentage of 0.5-1.0$\%$ \cite{ichimura2015conduction}, where there is enough Cu present to modify the entire band structure. Extensive doping is limited by the low solubility of Cu in ZnS due to dissimilarity in the crystal structures of ZnS and Cu$_2$S. Consequently, Cu$_x$S phases are commonly observed as precipitates on the surface of films \cite{ichimura2015conduction}. A number of methods have recently been employed to synthesise ZnS:Cu, including sputtering \cite{chamorro2016role} and solution-based approaches \cite{ortiz2014p,dula2012photochemical}.

Diamond et al. recently synthesised ZnS:Cu thin films using pulsed laser deposition (PLD), and reported conductivities of 54.4 Scm$^{-1}$ and a transmission of 65$\%$ for films of approximately 100 nm thickness \cite{diamond2012copper}. Hole mobility and carrier density for high-temperature PLD-fabricated ZnS:Cu thin films remain unreported. More recently, Woods-Robinson et al. synthesised ZnS:Cu at low temperature (T $<$ 100$^\circ$C) also using PLD, achieving a maximum conductivity of 42 Scm$^{-1}$, and hole concentrations of 1-2 $\times$10$^{20}$cm$^{-3}$ in films with a Cu content of 30$\%$ \cite{woods2016p}. Noting the relatively high dopant concentration, the small free carrier concentrations suggest partial impurity-related compensation. Our preliminary work found that post-deposition annealing in vacuum can render sulphide thin films S-deficient, with increased S vacancies resulting in high levels of compensation. Despite the noted problems, it is evident that ZnS:Cu has the potential to be a versatile and useful p-type TCM for use in opto-electronics. 

This paper is organised as follows. In Sec. \ref{methods}, the experimental details of the fabrication and characterisation are provided. Sec. \ref{results} presents an optical and morphological characterisation of zincblende pulsed-laser-deposited ZnS:Cu before exploring the opto-electrical properties of the films and the effects of doping on hole mobility and carrier density. The paper concludes with a summary in Sec. \ref{conclusion}.

\section{Experimental Details}
\label{methods}
ZnS:Cu thin films were grown on fused quartz substrates (10 x 10 mm$^2$) by PLD, using an ArF UV excimer laser (coherent, model COMPexPRO 102 F-Version), with wavelength $\lambda$ = 193 nm. In-house ceramic targets comprised of 99.9$\%$ purity ZnS:Cu$_2$S were used, and a deposition chamber temperature of 450$^\circ$C was maintained throughout the depositions. Prior to film fabrication, the deposition chamber was evacuated to a pressure below 10$^{-5}$ Torr by a rotary vacuum pump, after which argon gas was introduced at a pressure of 30 mTorr. The substrate-target distance was 45 mm. Full deposition parameters are given in Table \ref{tab1}. Post-deposition, the samples underwent rapid thermal annealing (RTA) inside the vacuum chamber. The annealing process lasted 15 minutes, during which the films were raised to a temperature of 550$^\circ$C in a background gas of H$_2$S, before being rapidly brought down to room temperature. 

\begin{table}[b]
\centering
\begin{tabular}{lll}
\hline
Parameter                       &  & Value \\ \hline
Argon pressure/mTorr            &  & 30     \\
Laser pulse frequency/Hz        &  & 10     \\
Laser fluence/Jcm$^{-2}$        &  & 1.6    \\
Substrate temperature/$^\circ$C &  & 450    \\
Substrate-target distance/mm    &  & 45     \\ \hline
\end{tabular}
\caption{Experimental deposition parameters for the pulsed-laser deposition of the Cu-doped ZnS films.}\label{tab1}
\end{table}

The morphology of the samples was investigated by atomic force microscopy (AFM) and scanning electron microscopy (SEM, HITACHI S2300), and the microstructure was determined from X-ray diffraction analysis (XRD, PANalytical X'Pert PRO) using a Cu K$_\alpha$ source. The film composition and thicknesses were found with energy-dispersive X-ray analysis and focused ion beam spectroscopy (EDX and FIB, FEI Helios NanoLab 600 DualBeam). Hall-effect measurements were carried out to determine the conductivity, carrier mobility, and carrier concentration, using the van der Pauw method. The optical properties of the films were obtained with ultraviolet-visible spectroscopy (UV-Vis, Shimadzu UV-2600).

\section{Results and Discussion}
\label{results}

\subsection{Structural and morphological characterisation}

\begin{figure}[h]
\setlength{\abovecaptionskip}{0pt}
\setlength{\belowcaptionskip}{10pt}
\includegraphics[width=0.85\linewidth]{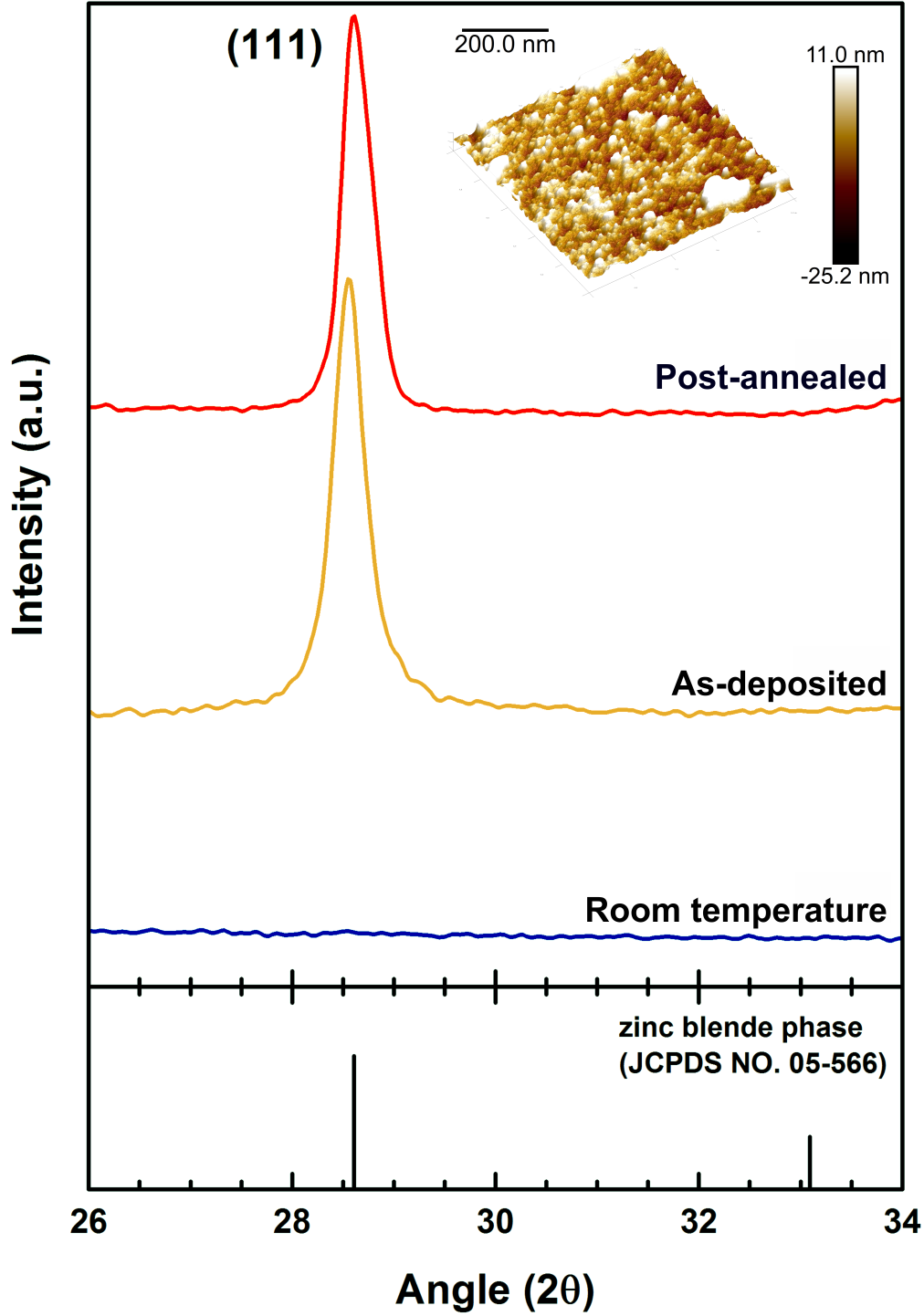}
\caption{XRD spectra plotted as a function of Cu K$\alpha$ 2$\theta$ diffraction angle. Diffraction standards of the zincblende crystal phase are represented by the vertical lines in the lower subplot. Top: XRD spectrum of a post-annealed Cu$_{0.33}$Zn$_{0.67}$S film displaying the characteristic (111) peak of zincblende ZnS. Middle: XRD spectrum of an as-deposited Cu$_{0.22}$Zn$_{0.78}$S film. Bottom: XRD pattern of a thin film deposited at room temperature. Inset: AFM image of a Cu$_{0.25}$Zn$_{0.75}$S film deposited under the rotation setup. } \label{fig1}
\end{figure}

The $\theta$-2$\theta$ XRD patterns of the ZnS:Cu thin films were recorded in the range 2$\theta$ = 25$^\circ$-45$^\circ$, and a representative set of data are presented in Fig. \ref{fig1}. The results indicate that films deposited at 450$^\circ$C have the polycrystalline structure, while room-temperature depositions lead to amorphous ZnS:Cu. Diffraction peaks appeared at the 2$\theta$ value of 29$^\circ$, corresponding to reflections from the the (111) lattice plane of the zincblende phase of ZnS. The crystal structure of the films is in good agreement with (JCPDS 05-566) zincblende ZnS data, with no trace of any minority wurtzite phase in either the as-deposited films or the annealed samples. No significant shift for the doped ZnS:Cu film is observed in the (111) peak in comparison to the undoped ZnS film, as can be expected from the similar crystal radii of Cu and Zn.

\begin{figure*}[]
\vspace{5mm} 
\includegraphics[width=0.7\linewidth]{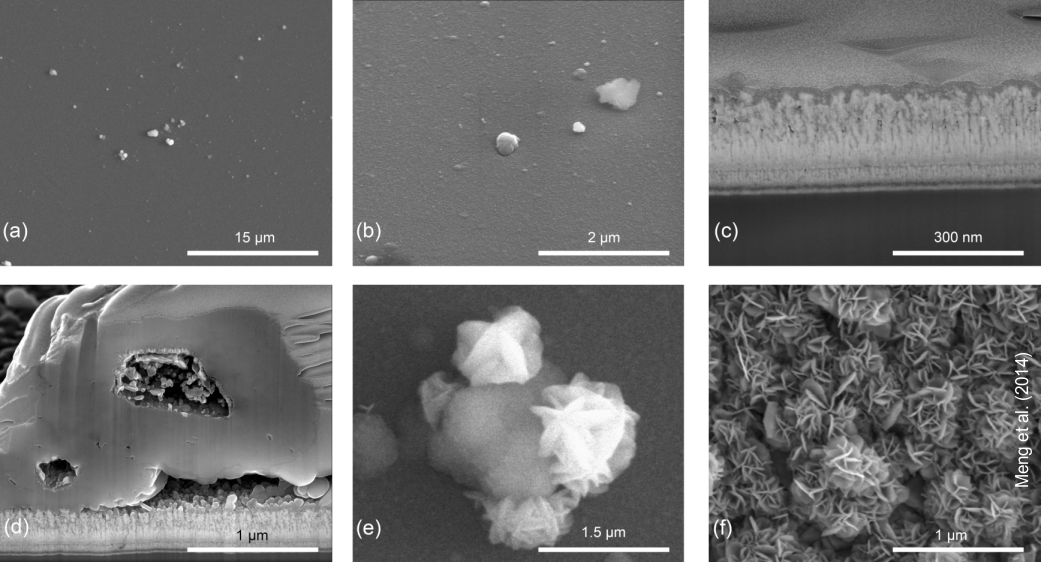}
\caption{SEM images of film surfaces: (a) 5000x magnification of film surface showing Cu$_2$S precipitates; (b) 35,000x magnification of film surface showing surface contaminants and overall uniformity; (c) Cross-sectional film displaying clear columnar growth; (d) Particulate deposited on a substrate during a deposition with a low-density target; (e) Image of Cu$_2$S nanoflowers on the surface of a film post-deposition; (f) Cu$_2$S nanoflowers synthesised by Meng et al. for comparison \cite{meng2014efficient}.} \label{fig2}
\end{figure*}

\begin{table*}[]
\centering
\begin{tabular}{llllll}
\hline
Sample ID & (hkl)        & FWHM (degrees) & D (nm) & $\delta$$\times$10$^{-3}$ (nm$^{-2}$) & $\epsilon$$\times$10$^{-2}$ \\ \hline
CuZnS\_1  & (111)$\beta$ & 0.33543        & 24.17  & 1.712                                          & 1.483                \\
CuZnS\_2  & (111)$\beta$ & 0.11265        & 71.95  & 0.193                                          & 0.331                \\
CuZnS\_3  & (111)$\beta$ & 0.36704        & 22.09  & 2.049                                          & 1.261                \\
CuZnS\_4  & (111)$\beta$ & 0.36379        & 22.30  & 2.011                                          & 1.250           \\ \hline    
\end{tabular}
\caption{Estimated FWHM, crystallite size D, dislocation density $\delta$ and strain $\epsilon$ of conducting ZnS:Cu films.}\label{tab2}
\end{table*}

Using the Scherrer equation, the lower bound on crystallite size was found to vary in the range 22-72 nm, with the size appearing to depend heavily on the deposition conditions to such an extent that correlations with both Cu content and post-deposition annealing were not apparent. In addition to the grain size, the dislocation density and strain of the films were obtained using the Scherrer method on the dominant peak in order to compare the data with literature values. These results are presented in Table \ref{tab2} and are found to resemble those in previous studies \cite{yildirim2012effect,yildirim2009annealing}. The films were not found to be subject to any considerable strain, a result which is further corroborated by a lack of film cracking, as seen in Fig. \ref{fig2}. 

The surface morphology of the as-grown ZnS:Cu films was characterized by SEM and FIB. The films display clear columnar growth, with dense structures possessing a high degree of binding within the columns and at the boundaries between them, and the formation of well-defined grain borders (Fig. \ref{fig2} (c)). Given that the normalised temperature during deposition was approximately 0.5, this structure is consistent with theory \cite{yildirim2009annealing,petrov2003microstructural}. AFM showed approximate crystallite size in the range of 10-20 nm (Fig. \ref{fig1}, inset). This is in accordance with the XRD findings. Surface spots, examples of which are present in Fig. \ref{fig2} (a,b) were found by EDX measurement and SEM to be Cu$_2$S surface defects. This is expected, as Cu is only soluble in ZnS up to a concentration of around 400 ppm, and upon reaching this limit, Cu is often found near the surface of the ZnS films or in phase-separated Cu$_x$S precipitates \cite{pham2015dft+,ichimura2015conduction}. After annealing, these precipitates became sparser, and it was observed that they displayed the characteristic structure of Cu$_2$S nanoflowers (Fig. \ref{fig2} (e)). Aside from these defects, the film surfaces possessed high uniformity and appeared to be continuous, dense, and well-adhered to the substrate. FIB measurements determined the thicknesses of the films to be 100$\pm$5 nm.

\subsection{Composition and elemental distribution}

\begin{figure}[]
\setlength{\abovecaptionskip}{0pt}
\setlength{\belowcaptionskip}{10pt}
\includegraphics[width=\linewidth]{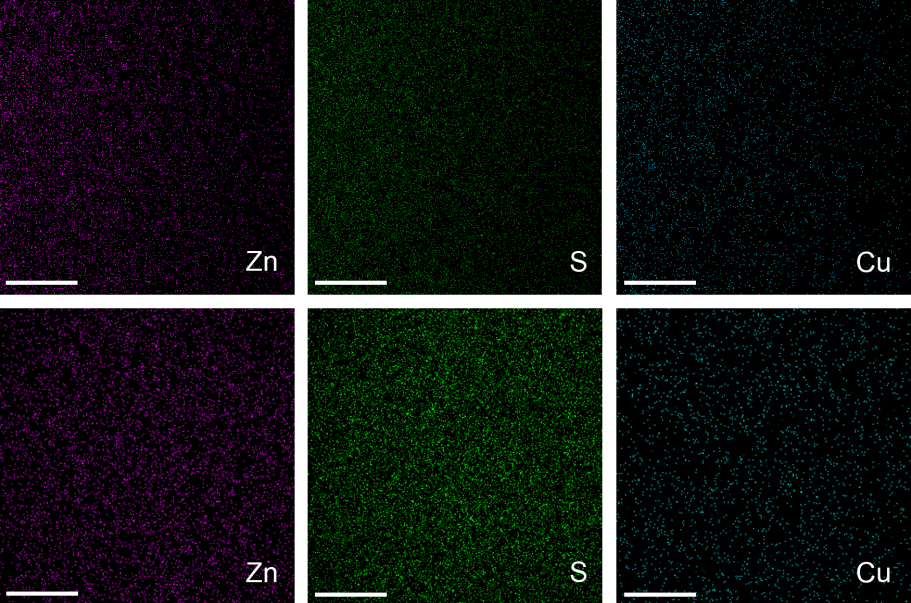}
\vspace{1mm}
\caption{\label{fig3}Elemental mapping showing the distribution of deposited elements in representative conducting ZnS:Cu films. Mapping was performed on a central area of the films. Top: elemental maps of a stationary-substrate-grown film. Note the non-uniform distribution from left to right in each elemental map. Bottom: elemental maps for a film grown under the rotation setup, showing high levels of uniformity across the film.} 
\end{figure}

Quantitative and semi-quantitative EDX analysis was performed on several films in order to determine stoichiometry and the distribution of the constituent elements. The first set of fabricated films were deposited on stationary substrates. The geometry of the setup and the plume dynamics resulted in the substrates being subject to differential stoichiometry across their areas, and the consequent growth was found to be highly sensitive to initial conditions (Fig. \ref{fig3}). The atomic percentages of the more volatile element, S, were found to be relatively stable across the films, however the films often possessed many atomic-percent of non-stoichiometry with respect to the heavier, more directional, Cu and Zn cations. In contrast, films deposited under the rotating setup were found to have high uniformity of elemental distribution, even when stoichiometry with respect to the target material was low (see Fig. \ref{fig3} (c-e)). Post-deposition annealing in a S-rich background gas prevented sulphur vacancy (V$_S$) formation during the heating process and otherwise kept the stoichiometry stable.

\subsection{Electrical characterisation}

\begin{figure*}[]
\vspace{5mm} 
\includegraphics[width=\linewidth]{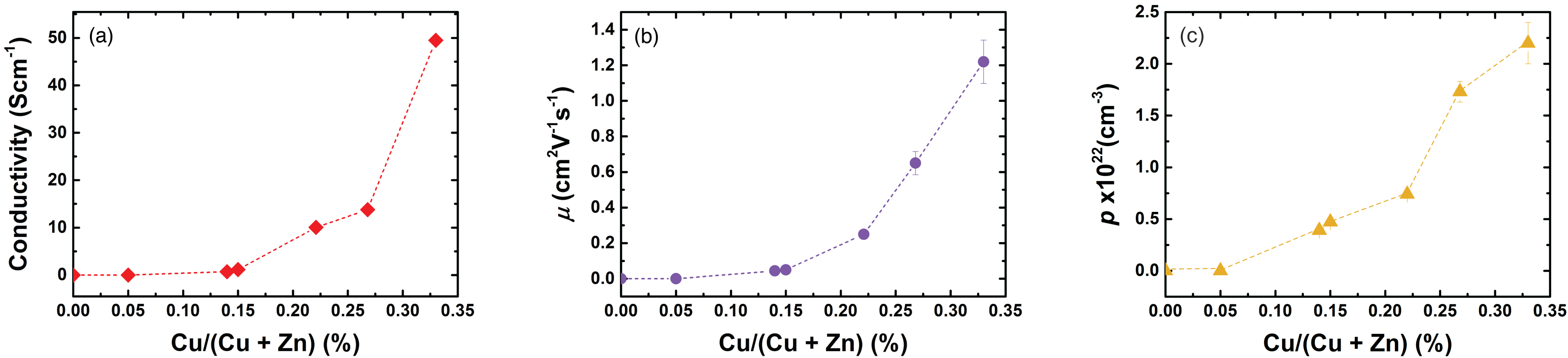}
\caption{\label{fig4}Electrical properties at room temperature plotted as a function of x. Relation between copper content and conductivity (a), hole mobility (b), and free carrier concentration (c). A maximum conductivity of 49.0 Scm$^{-1}$ was found at x = 0.33 which corresponds to a hole mobility of 1.22 cm$^2$V$^{-1}$s$^{-1}$.} 
\end{figure*}

P-type behaviour of the ZnS:Cu films was confirmed by Hall measurement from which sheet resistance, hole concentration, and carrier mobility in the annealed samples were calculated. The highest recorded mobility of 49 Scm$^{-1}$ was found at a Cu concentration of x = 0.33 (Fig. \ref{fig4}). Consistent with previous studies, it was found that an increase in conductivity was correlated with an increasing Cu concentration. For values of x = 0.05 and below, the films were found to be non-conducting. A significant barrier to high conductivity is the compensatory action of defect formation. As a result of compensating V$_S$ native defects, ZnS cannot be naturally p-type, and therefore extrinsic Cu-acceptors are the main source of free carriers \cite{li2012native}. The total hole concentration in ZnS:Cu therefore depends both on the extent of defect formation and on the number of Cu ions that successfully substitute for Zn ions in the ZnS lattice. At copper concentrations of x = 0.10 and above, the films experience sufficient cation substitution to display p-type conducting behaviour. Conductivity, hole mobility, and carrier density are plotted as a function of copper concentration in Fig. \ref{fig4}, and a full electrical characterisation of the films is presented in Table \ref{tab3}. The hole mobility in the peak conductivity films was found to be 1.22 cm$^2$V$^{-1}$s$^{-1}$ with a corresponding free carrier density of 2.20 $\times$10$^{22}$ cm$^{-3}$. 

\begin{table*}[]
\vspace{5mm}
\centering
\begin{tabular}{lllll}
\hline
Cu$_x$Zn$_{1-x}$S & R$_S$ (k$\Omega$/$\Box$) & $\sigma$ (Scm$^{-1}$) & $\mu$ (cm$^2$V$^{-1}$s$^{-1}$) & p$_S$ ($\times$10$^{22}$ cm$^{-3}$) \\ \hline
x = 0             & -                 & -        & -                                                                           & -                          \\
x = 0.05          & -                 & -        & -                                                                           & -                          \\
x = 0.14          & 30.3              & 0.21     & 0.04                                                                        & 0.39                       \\
x = 0.15          & 19.0              & 0.65     & 0.05                                                                        & 0.42 \\
x = 0.22          & 9.70              & 9.54     & 0.25                                                                        & 0.74                       \\
x = 0.27          & 6.97              & 13.3     & 0.65                                                                        & 1.7                       \\
x = 0.33          & 7.80              & 49.0     & 1.22                                                                        & 2.2                     \\ \hline
\end{tabular}
\caption{\label{tab3}Summary of the electrical properties of a selection of ZnS:Cu films grown at various Cu concentrations with an average thickness of 105 nm.}
\end{table*}

\begin{figure}[]
\vspace{5mm}
\setlength{\abovecaptionskip}{0pt}
\setlength{\belowcaptionskip}{10pt}
\includegraphics[width=0.6\linewidth]{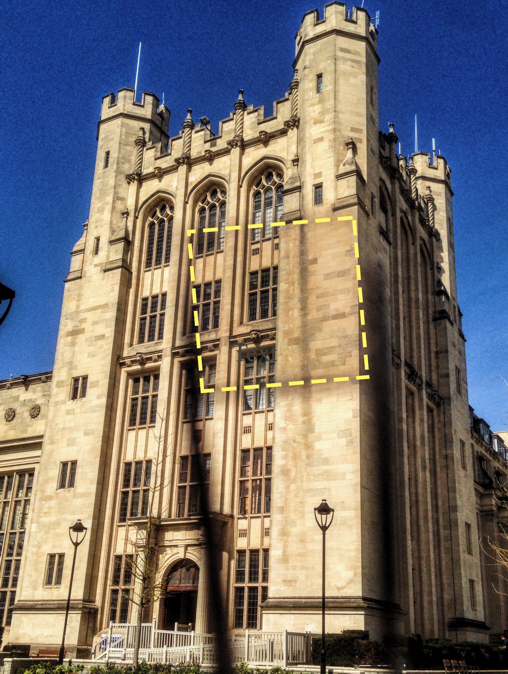}
\caption{Photograph demonstrating the high transparency of a typical film of 100 nm thickness.} \label{fig5}
\end{figure}
These data are consistent with the XRD findings, which show evidence of some secondary phase Cu$_2$S. Given the seeming surface localisation and relative sparsity of the Cu$_2$S (see Fig. \ref{fig2} (a,b)), however, we instead believe that ineffective doping of Cu within the film is the origin of these defect states. The lack of sub-band gap absorption in the rotating-substrate films gives further weight to this hypothesis. 

\begin{figure*}[]
\setlength{\abovecaptionskip}{0pt}
\setlength{\belowcaptionskip}{10pt}
\includegraphics[width=0.8\linewidth]{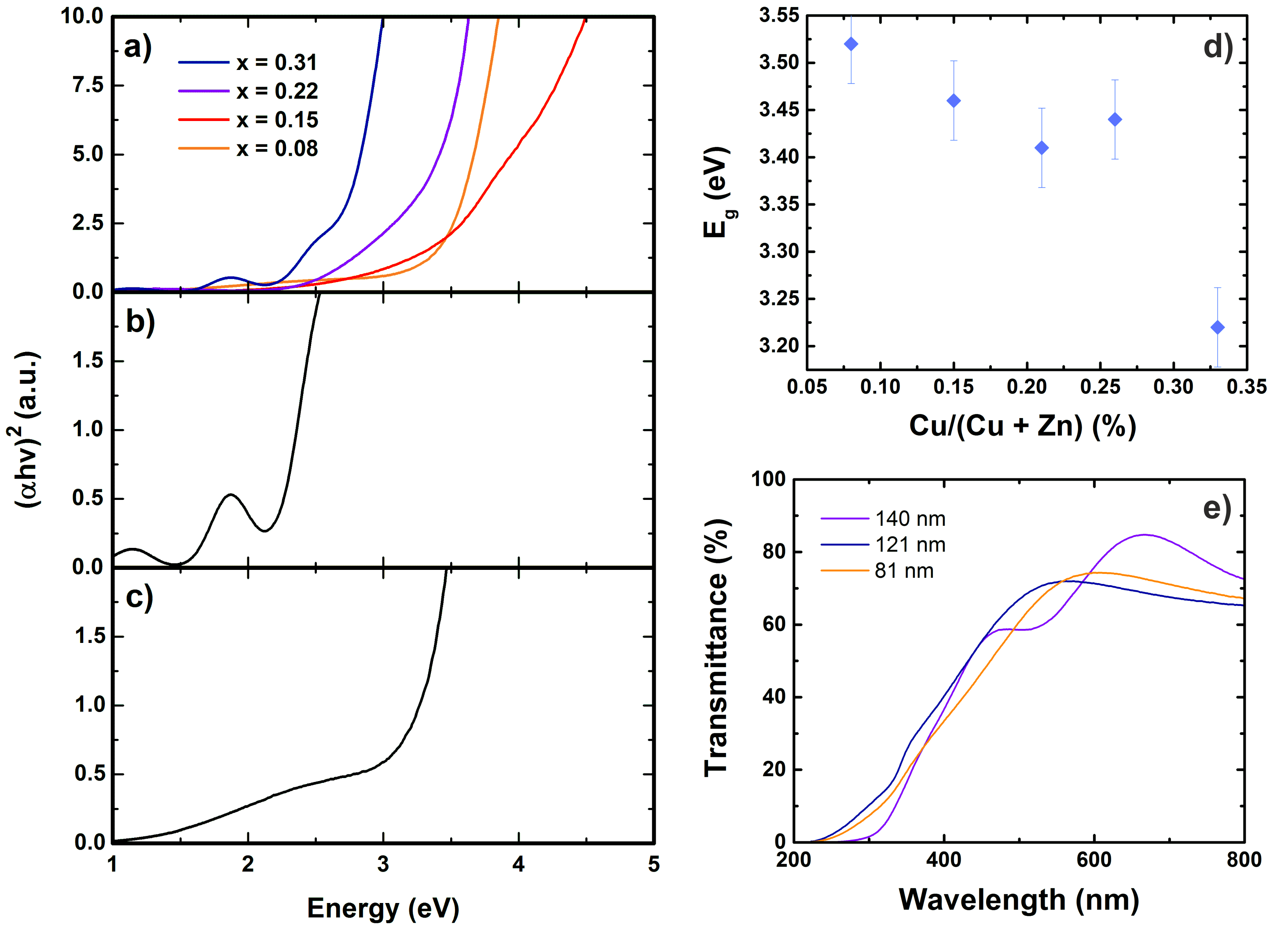}
\caption{Tauc plots derived from transmittance spectra. (a) Tauc plot of four Cu$_x$Zn$_{1-x}$S films showing the reduction in absorption onset with an increase in copper concentration. (b) Tauc plot of a high Cu-content film grown on a stationary substrate. (c) Tauc plot of a low Cu-content film grown on a stationary substrate. (d) Optical band gap plotted as a function of copper concentration. The band gaps are found to vary between 3.2-3.53 eV. (e) Transmittance spectra for Cu$_{0.33}$Zn$_{0.67}$S films of approximately 100 nm thickness grown on rotated substrates. Optical transmission at 550 nm displays an appreciable decrease at 140 nm thickness from 71.8$\%$ to 64.0$\%$.} \label{fig6}
\end{figure*}

Post-deposition, the films underwent RTA. If the annealing process leads to S-poor conditions, spontaneous formation of zinc interstitials (Zn$_i$) will compensate free holes, making p-type doping unobtainable. Under S-rich conditions, the formation energy of this defect increases significantly, and thus copper doping should lead to p-type behaviour. The RTA was therefore performed in the presence of H$_2$S to reduce S-outgassing. The conductivity of the films was found to have increased post-deposition annealing by a factor of 1.88$\pm$0.8, with the effect being most prominent in initially low conductivity films. This is hypothesised to correspond predominantly to the annealing-induced increase in crystallinity, given that the annealing had no major effect on optical transmission values. The results of Hall measurements on films deposited on stationary substrates demonstrated high levels of non-uniformity in conductivity across the surface, as effective Cu incorporation is highly dependent on plume geometry. Probed resistances between opposing diagonals in the least uniform non-rotating films displayed large discrepancies. Subjecting these films to a 15-minute RTA process rectified this situation, homogenizing the resistance across the film. We suggest that the annealing process encouraged Cu ions to substitute for Zn more evenly throughout the samples.

\subsection{Optical characterisation}

Transmittance measurements were carried out by UV-Vis spectroscopy at room temperature. Transparency of ZnS:Cu films of 100 nm thickness was found to reach a maximum of 82.6$\%$ in an x = 0.05 sample, with this value falling to 71.8$\%$ in the most conducting x = 0.33 films. (Fig. \ref{fig5}). 
The energy band gaps of Cu$_x$Zn$_{1-x}$S films were calculated by using optical transmittance spectra. To determine the energy band gap values, Tauc plots were produced, from which band gap energies could be determined by the extrapolation of the linear regions to the energy axis. Fig. \ref{fig6} (a) shows the shift in absorption onset with increasing Cu percentage to lower energies, which is indicative of the states introduced at the valence band maximum through Cu incorporation. For good optical transparency, ZnS:Cu must have a band gap of greater than 3 eV, so that only light of wavelengths shorter than the visible ($<$ 400 nm) can be absorbed by the material. In the ZnS:Cu films grown on stationary substrates, significant sub-band gap absorption was observed, as evidenced in Fig. \ref{fig6} (b). This could be indicative of valence band tailing or of the presence of a secondary phase, and occurs to a greater extent in Cu-rich films. This is also consistent with previous work displaying the reduction in doping efficiency with increasing dopant concentration \cite{byrappa2003crystal}. Fig \ref{fig6} (c) displays the reduction in this sub-band gap absorption behaviour with lowered Cu-content. 

Fig. \ref{fig6} (d) shows the smallest observed band gap for various Cu contents. Band gaps ranged from 3.20-3.53 eV, with a reduction from the zincblende ZnS value of 3.54 eV with increasing copper doping \cite{kumar1999band}. Fig. \ref{fig6} (e) shows typical transmittance spectra for three rotated ZnS:Cu films. The interference fringes in the spectra indicate that the films prepared under the rotating conditions display high uniformity. 

Zn$_i$, V$_S$, and Cu(I) formation are theorised to create deep energy levels with electron excitations that correspond to visible wavelengths, reducing transparency \cite{granqvist2002transparent}. It was observed that annealing did not affect transmission values within the visible and ultraviolet range, and had no significant effect on the band gap of the films, which suggests that sulphur outgassing and consequent vacancies were prevented during H$_2$S-rich annealing. 

\section{Summary}
\label{conclusion}
Optical and electrical properties of PLD-grown transparent p-type Cu-doped ZnS films were investigated and a comprehensive opto-electrical characterisation of the material is presented. Hall measurements have shown that ZnS:Cu thin films display electrical properties among the best reported for a p-type TCM, and UV-Vis spectroscopy has determined a high optical transmittance, with a band gap dependent on Cu-content. The most conducting Cu$_x$Zn$_{1-x}$S films possessed a conductivity of 49 Scm$^{-1}$ and a transmittance of 71.8$\%$ in the visible at x = 0.33.  A comparison of films grown on stationary and rotating substrates demonstrated the sensitivity of Cu-doping on the plume geometry, with the stationary films demonstrating differential conductivity across the films and sub-band gap absorption. We conclude that, for compositions of x = 0.10 and above, copper doping onto Zn sites is sufficient to overcome native compensation, and that Cu undergoes partial separation into Cu$_2$S precipitates. Application of the deposition method in the formation of SIS and MIS heterostructures constitutes a promising approach for the development of earth-abundant TCM-based opto-electronic devices.

\acknowledgments{We thank our colleagues Jonathan A. Jones and Sean A. Davis who helped in the collection of EDX data. Generous assistance was also provided by Dr. Peter Heard, Dr. Ross Springell, Dr. Robert Harniman and Dr. Benjamin Mills with the DualBeam, XRD, AFM, and UV-Vis equipment, respectively, throughout the course of the research. }

\bibliography{ZnS_Cu_APL}

\end{document}